\documentclass[sigconf]{acmart}

\usepackage{enumitem}
\usepackage{color, colortbl}

\AtBeginDocument{%
  \providecommand\BibTeX{{%
    \normalfont B\kern-0.5em{\scshape i\kern-0.25em b}\kern-0.8em\TeX}}}

\newcommand{\wh}{Wheeler}
\newcommand{\sysname}{Wheeler}
\newcommand{\wa}{\texttt{\small Wheel-1}}
\newcommand{\wb}{\texttt{\small Wheel-2}}
\newcommand{\wc}{\texttt{\small Wheel-3}}
\newcommand{\hnav}{\texttt{\small H-nav}}
\newcommand{\fnav}{\texttt{\small 2d-nav}} 
\newcommand{\tnav}{\texttt{\small 2d-T-nav}} 

\def \billahdebug{}

\ifx \billahdebug \undefined
\newcommand{\fixmesb}[1]{{}}
\newcommand{\fixmeth}[1]{{}}
\newcommand{\fixmens}[1]{{}}
\newcommand{\fixmera}[1]{{}}

\else

\newcommand{\up}[1]{{\bf\textcolor{blue}{~(\faThumbsUp)~}}}
\newcommand{\down}[1]{{\bf\textcolor{red}{~(\faThumbsDown)~}}}

\newcommand{\fixmesb}[1]{{\bf\textcolor{red}{ [ sB FIXME: #1 ]}}}
\newcommand{\fixmeth}[1]{{\bf\textcolor{blue}{ [ tH FIXME: #1 ]}}}

\fi
\newcolumntype{L}[1]{>{\raggedright\let\newline\\\arraybackslash\hspace{0pt}}m{#1}}
\newcolumntype{C}[1]{>{\centering\let\newline\\\arraybackslash\hspace{0pt}}m{#1}}
\newcolumntype{R}[1]{>{\raggedleft\let\newline\\\arraybackslash\hspace{0pt}}m{#1}}
\copyrightyear{2024}
\acmYear{2024}
\setcopyright{rightsretained}
\acmConference[UIST Adjunct '24]{The 37th Annual ACM Symposium on User Interface Software and Technology}{October 13--16, 2024}{Pittsburgh, PA, USA}
\acmBooktitle{The 37th Annual ACM Symposium on User Interface Software and Technology (UIST Adjunct '24), October 13--16, 2024, Pittsburgh, PA, USA}
\acmDOI{10.1145/3672539.3686749}
\acmISBN{979-8-4007-0718-6/24/10}

\begin{document}

\title[Demonstration of Wheeler]{Demonstration of Wheeler: A Three-Wheeled Input Device for Usable, Efficient, and Versatile Non-Visual Interaction}

\author[MT Islam]{Md Touhidul Islam$^*$}
\affiliation{%
  \institution{Pennsylvania State University}  
  \city{University Park}
  \state{PA}
  \country{USA}
}
\email{touhid@psu.edu}

\author[N Sojib]{Noushad Sojib$^*$}
\affiliation{%
  \institution{University of New Hampshire}  
  \city{Durham}
  \state{NH}
  \country{USA}
}
\email{noushad.sojib@unh.edu}

\author[I Kabir]{Imran Kabir}
\affiliation{%
  \institution{Pennsylvania State University}  
  \city{University Park}
  \state{PA}
  \country{USA}
}
\email{ibk5106@psu.edu}

\author[AR Amit]{Ashiqur Rahman Amit}
\affiliation{%
  \institution{Innovation Garage Limited}  
  \city{Dhaka}
  \country{Bangladesh}
}
\email{amit@innovationgarage.com.bd}

\author[M Ruhul Amin]{Mohammad Ruhul Amin}
\affiliation{%
  \institution{Fordham University}  
  \city{Bronx}
  \state{NY}
  \country{USA}
}
\email{mamin17@fordham.edu}

\author[SM Billah]{Syed Masum Billah}
\affiliation{%
  \institution{Pennsylvania State University}  
  \city{University Park}
  \state{PA}
  \country{United States}
}
\email{sbillah@psu.edu}

\thanks{$^*$Equal Contribution}

\begin{abstract}
Navigating multi-level menus with complex hierarchies remains a big challenge for blind and low-vision users, who predominantly use screen readers to interact with computers. To that end, we demonstrate Wheeler, a three-wheeled input device with two side buttons that can speed up complex multi-level hierarchy navigation in common applications. When in operation, the three wheels of Wheeler are each mapped to a different level in the application hierarchy. Each level can be independently traversed using its designated wheel, allowing users to navigate through multiple levels efficiently. Wheeler's three wheels can also be repurposed for other tasks such as 2D cursor manipulation. In this demonstration, we describe the different operation modes and usage of Wheeler.
\end{abstract}



\begin{teaserfigure}
    \centering
    \includegraphics[width=.99\textwidth]{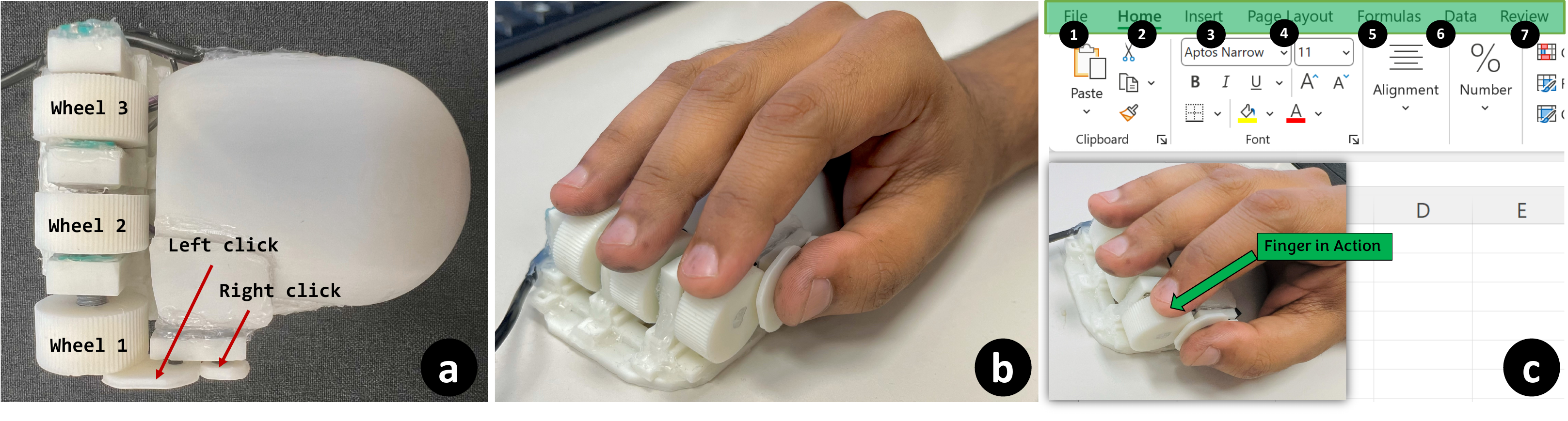}
    \caption{
     \textbf{{\sc \textbf{\sysname{}}} input device and its usage}.  (a) shows a 3D-printed implementation of \sysname{} having three wheels and two push buttons (primary and secondary) on the side;
    (b) shows a blind user holding the device, placing three central fingers on the three wheels, and the thumb over the two side buttons;
    (c) shows a user navigating the multi-level, hierarchical menu of Microsoft Excel's ribbon using \sysname{}'s \hnav{} mode.
    }
  \label{fig:teaser}
 \end{teaserfigure}

\maketitle
\section{Introduction}
\label{sec:intro}
Navigating the complex hierarchies in modern desktop applications remains one of the key accessibility challenges for blind and low-vision users, who use a combination of keyboard and screen readers (e.g., NVDA~\cite{NVAccess38:online}, JAWS~\cite{jaws_screenshade}, VoiceOver~\cite{voiceover_osx}) to interact with computers.
Recent studies have shown that apps requiring a higher average number of keystrokes for navigation are perceived as less accessible~\cite{touhidul2023probabilistic}.
As such, developing a faster mechanism to travel between menu items in an app is necessary.

While prior research has proposed alternate input modalities with faster task completion times in specific scenarios~\cite{Billah_speeddial, haena-speeddial2, haena-speeddial1}, the challenge of navigating to UI items that belong to a different sub-tree remains.
To address this, we design and implement \textsc{\textbf{\sysname{}}}~\cite{islam2024wheeler}, a three-wheeled, mouse-shaped, stationary input device whose three wheels can be mapped to three different levels in an app's internal hierarchy---enabling faster and more convenient navigation.
The three wheels also offer versatility such as the ability to manipulate cursor in 2D space.

\section{\sysname{}: an Overview}
\sysname{} is a mouse-shaped input device with three wheels and two side push buttons, as shown in Figure~\ref{fig:teaser}a.
Unlike a mouse, \sysname{} is stationary, i.e.,  users do not move it on the surface when using it. 
A user can grip the device with their right hand so that their index finger rests on the first wheel, 
the middle finger on the second wheel, the ring finger on the third wheel, and the thumb over the two buttons as shown in Figure~\ref{fig:teaser}b.
Of the two side buttons, the bigger one plays the role of a mouse left/primary click, and the smaller one plays the role of a mouse right/ secondary click.

In our design, \sysname{} connects to a computer via USB, but a Bluetooth wireless connection is feasible.
\sysname{} provides audio-haptic feedback to convey cursor context.
It has a buzzer and haptic motor; the buzzer beeps during significant events, and the haptic motor vibrates with each rotation. These do not interfere with screen reader audio.

\section{Interaction Using \sysname{}}
\sysname{} primarily operates in two modes: \textbf{\hnav{}} and \textbf{\fnav{}}.

\begin{figure}[!ht]
    \centering
    \includegraphics[width=\columnwidth]{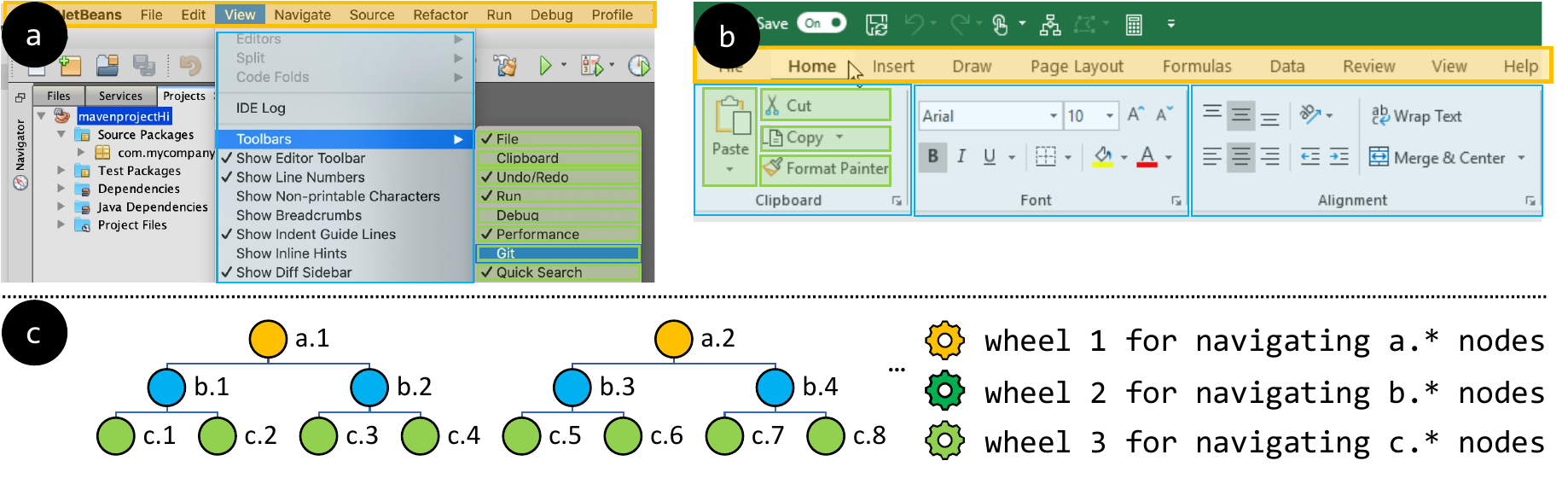}
    \caption{\textbf{{\sysname{}'s 
    \hnav{} mode in action.
    (a) and (b) shows the three-level hierarchies of two applications.
    Yellow, blue, and green colors represent the first, second, and third levels, respectively, in the hierarchy of either app.
    (c) shows the tree representation of either hierarchy---different levels of which can be traversed efficiently using \sysname{}'s wheels.}}}
    \label{fig:h-nav-demo}
\end{figure}   

\textit{\textbf{\hnav{} Mode.}}
In \hnav{} mode, \sysname{} navigates an app's abstract UI tree (Figure~\ref{fig:h-nav-demo}). 
By default, \sysname{}'s three wheels point to the top three levels of an app's DOM, each with its own cursor and state.
A wheel remembers the last UI object focused on and resumes from there, eliminating the need to re-explore the hierarchy.

The rotate action selects elements bi-directionally. 
\wa{} selects elements in the $1^{st}$ level, \wb{} selects children of \wa{}'s selection, and \wc{} selects children of \wb{}'s selection. 
When \wa{}'s cursor moves to a node, \wb{}'s cursor moves to the first child of \wa{}'s node, and \wc{}'s cursor moves to the first child of \wb{}'s node. 
Figure~\ref{fig:h-nav-demo}c shows the hierarchical organization and mapping in \sysname{}.

For left-/right-clicks, the user presses the primary/secondary side buttons. 
Users can define rotation resolution (degrees) to adjust sensitivity. 
\sysname{} provides audio-haptic feedback for valid operations and spatial information.

In \hnav{} mode, \sysname{}'s firmware integrates with NVDA, an open-source screen reader, appearing as an NVDA plugin~\cite{momotaz_plugin} with access to any app's UI hierarchy via NVDA APIs, which uses Windows' native UI Automation API~\cite{ui_automation} to extract the UI tree and relay rotational input.

\textit{Traversing Apps with More than 3 Levels.}
For applications with more than 3 levels, users can move all three cursors down one level in the hierarchy by holding the \texttt{CTRL} key and pressing \sysname{}'s primary button. 
Similarly, to move all three cursors up one level, users can hold the \texttt{CTRL} key and press \sysname{}'s secondary button.

\begin{figure}[!ht]
    \centering
    \includegraphics[width=0.75\columnwidth]{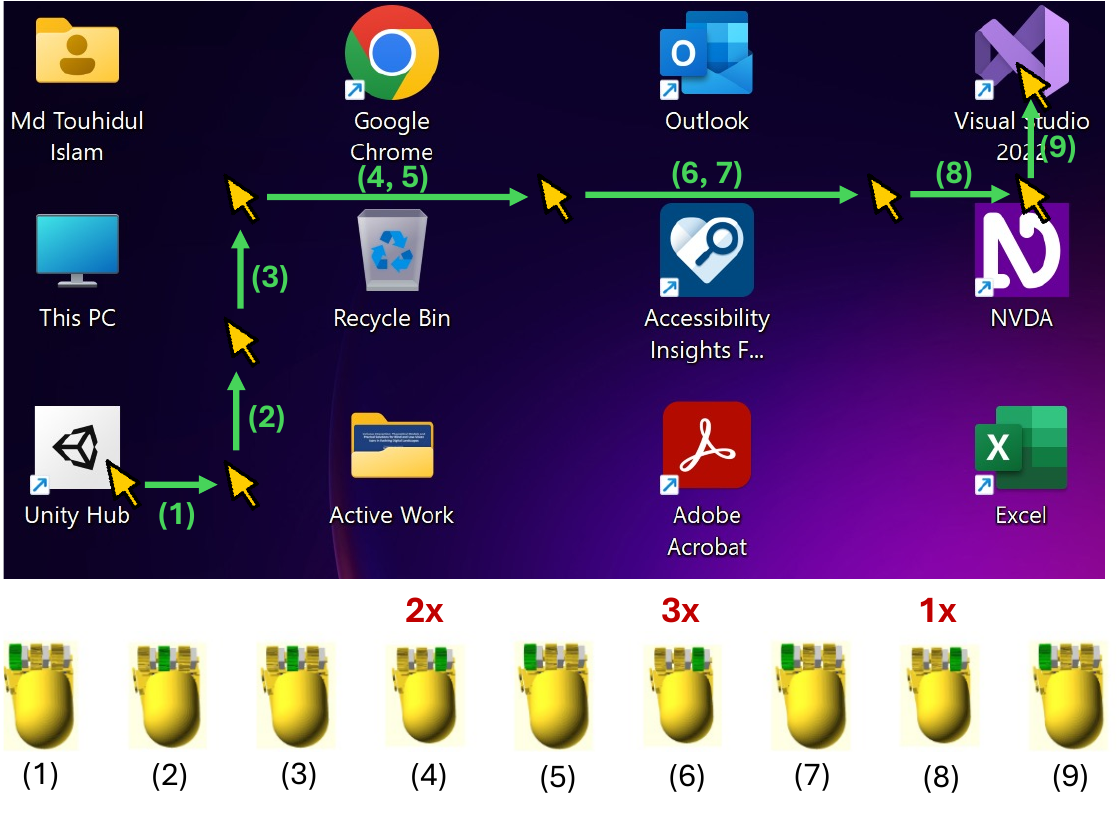}
    \caption{\textbf{{\sysname{}'s 
        \fnav{} mode in action.
        (Top) shows the steps in moving the cursor from the lower-left corner to the upper-right corner of a 2D screen.
        (Bottom) shows the sequence of wheel operations required at different steps to achieve this goal.}}}
    \label{fig:2d-nav-demo}
\end{figure}

\textit{\textbf{\fnav{} Mode.}}
In \fnav{} mode, the wheels serve different roles: \wa{} moves the cursor along the \texttt{X-axis}, \wb{} moves it along the \texttt{Y-axis}, and \wc{} controls the cursor speed. Figure~\ref{fig:2d-nav-demo} demonstrates a blind user moving the cursor from the lower-left to the upper-right corner of a 2D screen. The user can rotate \wc{} to adjust cursor speed.

Navigating 2D space can cause context loss for visually impaired users~\cite{islam2023spacex}. 
To address this, pressing the \texttt{CTRL} key in \fnav{} mode prompts \sysname{} to read out the cursor location as a percentage of the screen's width and height.
For example, if the cursor is above the \texttt{``Google Chrome''} icon in Figure~\ref{fig:2d-nav-demo}, \sysname{} would announce something like \emph{``30\% from the left and 10\% from the top''}. 
Additionally, \wh{}'s built-in TTS engine automatically reads out the name of a UI element on cursor hover.

\textit{\tnav{} Mode.}
\tnav{} is a variant of \fnav{} mode in which \sysname{} teleports the mouse cursor to the nearest neighboring UI element in the direction of cursor movement. 
This method is faster than \fnav{} for moving between elements.

\textit{\textbf{Toggling Modes.}}
To toggle between \hnav{} and \fnav{} modes, users can hold the \texttt{CTRL} button and simultaneously press both the primary and secondary buttons of \sysname{}. 
When in \fnav{} mode, users can enable or disable \tnav{} mode by pressing and holding the secondary (small) button for a short duration (e.g., 300 ms).

\bibliographystyle{ACM-Reference-Format}
\bibliography{Bibliography, Bibliography2, Bibliography3}


\begin{thebibliography}{11}


\ifx \showCODEN    \undefined \def \showCODEN     #1{\unskip}     \fi
\ifx \showDOI      \undefined \def \showDOI       #1{#1}\fi
\ifx \showISBNx    \undefined \def \showISBNx     #1{\unskip}     \fi
\ifx \showISBNxiii \undefined \def \showISBNxiii  #1{\unskip}     \fi
\ifx \showISSN     \undefined \def \showISSN      #1{\unskip}     \fi
\ifx \showLCCN     \undefined \def \showLCCN      #1{\unskip}     \fi
\ifx \shownote     \undefined \def \shownote      #1{#1}          \fi
\ifx \showarticletitle \undefined \def \showarticletitle #1{#1}   \fi
\ifx \showURL      \undefined \def \showURL       {\relax}        \fi
\providecommand\bibfield[2]{#2}
\providecommand\bibinfo[2]{#2}
\providecommand\natexlab[1]{#1}
\providecommand\showeprint[2][]{arXiv:#2}

\bibitem[\protect\citeauthoryear{??}{jaw}{2018}]%
        {jaws_screenshade}
 \bibinfo{year}{2018}\natexlab{}.
\newblock \bibinfo{title}{What's New in JAWS 2018 Screen Reading Software}.
\newblock
\newblock
\urldef\tempurl%
\url{https://www.freedomscientific.com/downloads/JAWS/JAWSWhatsNew}
\showURL{%
Retrieved September 19, 2018 from \tempurl}


\bibitem[\protect\citeauthoryear{??}{NVA}{2020}]%
        {NVAccess38:online}
 \bibinfo{year}{2020}\natexlab{}.
\newblock \bibinfo{title}{NV Access}.
\newblock \bibinfo{howpublished}{\url{https://www.nvaccess.org/}}.
\newblock
\newblock
\shownote{(Accessed on 09/20/2018).}


\bibitem[\protect\citeauthoryear{{Apple Inc.}}{{Apple Inc.}}{2020}]%
        {voiceover_osx}
\bibfield{author}{\bibinfo{person}{{Apple Inc.}}} \bibinfo{year}{2020}\natexlab{}.
\newblock \bibinfo{title}{VoiceOver}.
\newblock \bibinfo{howpublished}{\url{https://www.apple.com/accessibility/osx/voiceover/}}.
\newblock


\bibitem[\protect\citeauthoryear{Billah, Ashok, Porter, and Ramakrishnan}{Billah et~al\mbox{.}}{2017}]%
        {Billah_speeddial}
\bibfield{author}{\bibinfo{person}{Syed~Masum Billah}, \bibinfo{person}{Vikas Ashok}, \bibinfo{person}{Donald~E. Porter}, {and} \bibinfo{person}{I.V. Ramakrishnan}.} \bibinfo{year}{2017}\natexlab{}.
\newblock \showarticletitle{Speed-Dial: A Surrogate Mouse for Non-Visual Web Browsing}. In \bibinfo{booktitle}{\emph{Proceedings of the 19th International ACM SIGACCESS Conference on Computers and Accessibility}}. \bibinfo{publisher}{ACM}, \bibinfo{address}{3132531}, \bibinfo{pages}{110--119}.
\newblock
\showISBNx{ISBN}
\urldef\tempurl%
\url{https://doi.org/10.1145/3132525.3132531}
\showDOI{\tempurl}


\bibitem[\protect\citeauthoryear{Islam and Billah}{Islam and Billah}{2023}]%
        {islam2023spacex}
\bibfield{author}{\bibinfo{person}{Md~Touhidul Islam} {and} \bibinfo{person}{Syed~Masum Billah}.} \bibinfo{year}{2023}\natexlab{}.
\newblock \showarticletitle{SpaceX Mag: An Automatic, Scalable, and Rapid Space Compactor for Optimizing Smartphone App Interfaces for Low-Vision Users}.
\newblock \bibinfo{journal}{\emph{Proceedings of the ACM on Interactive, Mobile, Wearable and Ubiquitous Technologies}} \bibinfo{volume}{7}, \bibinfo{number}{2} (\bibinfo{year}{2023}), \bibinfo{pages}{1--36}.
\newblock


\bibitem[\protect\citeauthoryear{Islam, Donald, and Billah}{Islam et~al\mbox{.}}{2023}]%
        {touhidul2023probabilistic}
\bibfield{author}{\bibinfo{person}{Md~Touhidul Islam}, \bibinfo{person}{E~Porter Donald}, {and} \bibinfo{person}{Syed~Masum Billah}.} \bibinfo{year}{2023}\natexlab{}.
\newblock \showarticletitle{A Probabilistic Model and Metrics for Estimating Perceived Accessibility of Desktop Applications in Keystroke-Based Non-Visual Interactions}. In \bibinfo{booktitle}{\emph{The 2023 CHI Conference on Human Factors in Computing Systems (CHI '23)}}. \bibinfo{publisher}{ACM}.
\newblock
\urldef\tempurl%
\url{https://doi.org/10.1145/3544548.3581400}
\showDOI{\tempurl}


\bibitem[\protect\citeauthoryear{Islam, Sojib, Kabir, Amit, Ruhul~Amin, and Billah}{Islam et~al\mbox{.}}{2024}]%
        {islam2024wheeler}
\bibfield{author}{\bibinfo{person}{Md~Touhidul Islam}, \bibinfo{person}{Noushad Sojib}, \bibinfo{person}{Imran Kabir}, \bibinfo{person}{Ashiqur~Rahman Amit}, \bibinfo{person}{Mohammad Ruhul~Amin}, {and} \bibinfo{person}{Syed~Masum Billah}.} \bibinfo{year}{2024}\natexlab{}.
\newblock \showarticletitle{Wheeler: A Three-Wheeled Input Device for Usable, Efficient, and Versatile Non-Visual Interaction}. In \bibinfo{booktitle}{\emph{The 37th Annual ACM Symposium on User Interface Software and Technology}}. \bibinfo{publisher}{Association for Computing Machinery}, \bibinfo{address}{Pittsburgh, PA, USA}.
\newblock
\showISBNx{979-8-4007-0628-8/24/10}
\urldef\tempurl%
\url{https://doi.org/10.1145/3654777.3676396}
\showURL{%
\tempurl}


\bibitem[\protect\citeauthoryear{Lee, Ashok, and Ramakrishnan}{Lee et~al\mbox{.}}{2020a}]%
        {haena-speeddial2}
\bibfield{author}{\bibinfo{person}{Hae-Na Lee}, \bibinfo{person}{Vikas Ashok}, {and} \bibinfo{person}{I.~V. Ramakrishnan}.} \bibinfo{year}{2020}\natexlab{a}.
\newblock \showarticletitle{Repurposing Visual Input Modalities for Blind Users: A Case Study of Word Processors}. In \bibinfo{booktitle}{\emph{2020 IEEE International Conference on Systems, Man, and Cybernetics (SMC)}}. IEEE, \bibinfo{pages}{2714--2721}.
\newblock
\urldef\tempurl%
\url{https://doi.org/10.1109/SMC42975.2020.9283015}
\showDOI{\tempurl}


\bibitem[\protect\citeauthoryear{Lee, Ashok, and Ramakrishnan}{Lee et~al\mbox{.}}{2020b}]%
        {haena-speeddial1}
\bibfield{author}{\bibinfo{person}{Hae-Na Lee}, \bibinfo{person}{Vikas Ashok}, {and} \bibinfo{person}{I.~V. Ramakrishnan}.} \bibinfo{year}{2020}\natexlab{b}.
\newblock \showarticletitle{Rotate-and-Press: A Non-visual Alternative to Point-and-Click?}. In \bibinfo{booktitle}{\emph{HCI International 2020 -- Late Breaking Papers: Universal Access and Inclusive Design}}, \bibfield{editor}{\bibinfo{person}{Constantine Stephanidis}, \bibinfo{person}{Margherita Antona}, \bibinfo{person}{Qin Gao}, {and} \bibinfo{person}{Jia Zhou}} (Eds.). Springer, \bibinfo{publisher}{Springer International Publishing}, \bibinfo{address}{Cham}, \bibinfo{pages}{291--305}.
\newblock
\showISBNx{978-3-030-60149-2}


\bibitem[\protect\citeauthoryear{{Microsoft Inc.}}{{Microsoft Inc.}}{2020}]%
        {ui_automation}
\bibfield{author}{\bibinfo{person}{{Microsoft Inc.}}} \bibinfo{year}{2020}\natexlab{}.
\newblock \bibinfo{title}{{UI Automation Overview}}.
\newblock
\newblock
\urldef\tempurl%
\url{http://msdn.microsoft.com/en-us/library/ms747327.aspx}
\showURL{%
\tempurl}


\bibitem[\protect\citeauthoryear{Momotaz, Islam, Ehtesham-Ul-Haque, and Billah}{Momotaz et~al\mbox{.}}{2021}]%
        {momotaz_plugin}
\bibfield{author}{\bibinfo{person}{Farhani Momotaz}, \bibinfo{person}{Md~Touhidul Islam}, \bibinfo{person}{Md Ehtesham-Ul-Haque}, {and} \bibinfo{person}{Syed~Masum Billah}.} \bibinfo{year}{2021}\natexlab{}.
\newblock \showarticletitle{Understanding Screen Readers' Plugins}. In \bibinfo{booktitle}{\emph{The 23rd International ACM SIGACCESS Conference on Computers and Accessibility}}. \bibinfo{publisher}{ACM}, \bibinfo{pages}{1--10}.
\newblock
\urldef\tempurl%
\url{https://doi.org/10.1145/3441852.3471205}
\showDOI{\tempurl}


\end{thebibliography}

\appendix

\end{document}